\documentclass[conference]{IEEEtran}
\usepackage{threeparttable}
\usepackage{multirow}
\usepackage[caption=false]{subfig}
\usepackage{tabu}
\usepackage{color, colortbl}
\usepackage[linesnumbered,ruled,noend]{algorithm2e}
\usepackage{xcolor}
\usepackage{hyperref}
\usepackage{tabularx}
\usepackage{graphicx}
\usepackage[caption=false]{subfig}
\usepackage{threeparttable}
\def\BibTeX{{\rm B\kern-.05em{\sc i\kern-.025em b}\kern-.08em
    T\kern-.1667em\lower.7ex\hbox{E}\kern-.125emX}}
\usepackage{multirow}
\usepackage[caption=false]{subfig}
\usepackage{tabu}
\usepackage{color, colortbl}
\newcommand{\thickhline}{%
    \noalign {\ifnum 0=`}\fi \hrule height 1pt
    \futurelet \reserved@a \@xhline
}
\usepackage[linesnumbered,ruled,noend]{algorithm2e}
\usepackage{xcolor}

\SetCommentSty{mycommfont}    
\usepackage{pifont}
\usepackage{multirow}
\usepackage{hyperref}
\usepackage{tabularx}

\begin{document}

\title{Advancing Neuromorphic Computing: Mixed-Signal Design Techniques Leveraging Brain Code Units and Fundamental Code Units}

\makeatletter
\newcommand{\linebreakand}{%
  \end{@IEEEauthorhalign}
  \hfill\mbox{}\par
  \mbox{}\hfill\begin{@IEEEauthorhalign}
}
\makeatother

\author{%
\IEEEauthorblockN{Murat Isik\IEEEauthorrefmark{1}}
\IEEEauthorblockA{%
\textit{Stanford University}\\
Stanford, CA \\
Email: misik@stanford.edu\\
\textit{\IEEEauthorrefmark{1}Corresponding author}}
\and
\IEEEauthorblockN{Sols Miziev}
\IEEEauthorblockA{%
\textit{ni2o, Inc}\\
Providence, RI \\
sols.miziev@ni2o.com}
\and
\IEEEauthorblockN{Wiktoria Pawlak}
\IEEEauthorblockA{%
\textit{ni2o, Inc}\\
Providence, RI \\
wpawlak@ni2o.com}
\linebreakand 
\IEEEauthorblockN{Newton Howard}
\IEEEauthorblockA{%
\textit{University of Oxford}\\
Oxford, UK \\
newton.howard@brainsciences.org}
}
\thanks{\IEEEauthorrefmark{1}Corresponding author: Murat Isik, misik@stanford.edu}

\maketitle

\begin{abstract}
  This paper introduces a groundbreaking digital neuromorphic architecture that innovatively integrates Brain Code Unit (BCU) and Fundamental Code Unit (FCU) using mixed-signal design methodologies. Leveraging open-source datasets and the latest advances in materials science, our research focuses on enhancing the computational efficiency, accuracy, and adaptability of neuromorphic systems. The core of our approach lies in harmonizing the precision and scalability of digital systems with the robustness and energy efficiency of analog processing. Through experimentation, we demonstrate the effectiveness of our system across various metrics. The BCU achieved an accuracy of 88.0\% and a power efficiency of 20.0 GOP/s/W, while the FCU recorded an accuracy of 86.5\% and a power efficiency of 18.5 GOP/s/W. Our mixed-signal design approach significantly improved latency and throughput, achieving a latency as low as 0.75 ms and throughput up to 213 TOP/s. These results firmly establish the potential of our architecture in neuromorphic computing, providing a solid foundation for future developments in this domain. Our study underscores the feasibility of mixed-signal neuromorphic systems and their promise in advancing the field, particularly in applications requiring high efficiency and adaptability.
\end{abstract}

\section{Introduction}
Neuromorphic computing, a paradigm inspired by the neural structures and computational processes of the human brain, has seen considerable evolution since its inception. Tracing its origins to Hebb's concept of synaptic plasticity as a mechanism for learning and memory, it forms the foundation of modern neuromorphic computing \cite{Hebb1988}. The term "neuromorphic computing" was coined by Carver Mead in the late 1980s, marking a significant departure from traditional von Neumann architecture and leading to the development of the first neural-inspired chips like artificial retinas and cochleas using analog VLSI circuits \cite{MeadConway1978, MeadIsmail1989}.

In recent years, neuromorphic computing has expanded to include various implementations in software and hardware, including digital, analog, and mixed-signal circuits. This expansion is propelled by substantial research initiatives, such as the DARPA SyNAPSE program, fostering advancements in memristors, silicon neurons, and synapse models. Key developments in neuromorphic architectures include IBM's TrueNorth and Intel's Loihi, showcasing principles of plasticity and learning \cite{Hylton2008, Yin2017, Markram2012, DoE2020, Merolla2014, Davies2018}. Additionally, there is a growing exploration of optical "memristors" as potential key components for developing high-bandwidth and efficient neuromorphic machine learning hardware, which represents a recent and significant advancement in the field \cite{duan2024memristor, isik2023energy, isik2024neurosec}.

\subsection{Advantages of Neuromorphic Computing}

Neuromorphic systems excel in parallel processing, event-driven computation, and exhibit synaptic plasticity mechanisms like spike-timing-dependent plasticity (STDP), enhancing pattern recognition and sensory data processing. These systems are scalable and incorporate elements of stochasticity, reflecting the probabilistic nature of biological neural networks \cite{Tang2023, Ji2023, Aboumerhi2023, Bill2014, Buchel2022, Zhu2021, Furber2014, Davies2018}.

Brain Code Unit (BCU) and Fundamental Code Unit (FCU) are neuromorphic core of our proposed neuromorphic computing implementation. BCU is designed to mimic the brain's ability to encode and process information, while FCU serve as the fundamental building blocks for these operations. Together, they form the basis of neuromorphic systems, enabling them to replicate complex neural processes. Understanding these units is crucial for advancing neuromorphic computing and utilizing its full potential \cite{HowardHussain2018}.

\begin{figure}[h]
    \centering
    \includegraphics[width = 0.5\textwidth]{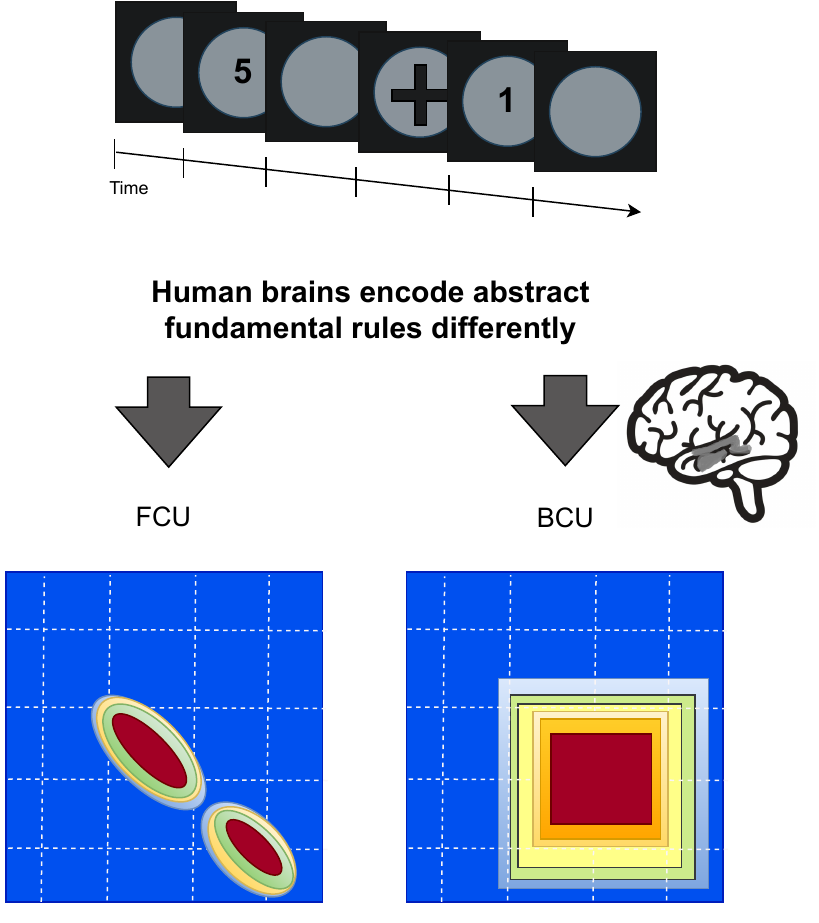}
    \caption{Differential Encoding of Abstract Mathematical Rules in Human. This figure illustrates the contrast between dynamic coding in the parahippocampal cortex and static coding in the hippocampus. The top panel depicts the sequence of abstract operations over time, linked to distinct neural activities. On the bottom left, dynamic coding is visualized as shifting activity patterns corresponding to different abstract rules. Conversely, the bottom right heatmap demonstrates static coding with consistent activation regions irrespective of changes in abstract rules. This indicates a division of labor within the human medial temporal lobe in the processing of abstract information, with implications for understanding the neural basis of high-level cognition.}
    \label{fig:neural_encoding}
\end{figure}

Mixed-signal design, which involves the integration of both analog and digital circuitry, plays a crucial role in the advancement of neuromorphic systems. This approach leverages the precision and flexibility of digital systems along with the robustness and energy efficiency of analog systems. Mixed-signal design holds the potential to significantly enhance the performance of neuromorphic computing architectures, particularly in terms of scalability, power efficiency, and the ability to handle a wide range of computational tasks.



\section{Background}\label{sec:nhw}
The landscape of neuromorphic computing is diverse, encompassing both digital and analog architectures. Digital neuromorphic systems, known for their precision and programmability, have been extensively explored and applied in various computational tasks. They simulate neural processes using discrete digital signals, offering a high degree of control and reproducibility. On the other hand, analog neuromorphic architectures attempt to more closely mimic the analog nature of biological neural networks. These systems are characterized by their energy efficiency and real-time processing capabilities, making them particularly suited for tasks that require natural, continuous data processing \cite{huynh2022implementing, isik2022design, isik2023survey, isik2023astrocyte}.

\subsection{Brain Code Unit (BCU) and Fundamental Code Unit (FCU)}

The development of Brain Code Unit (BCU) and Fundamental Code Unit (FCU) proposed by Newton Howard and Amir Hussain represents a significant advancement in neuromorphic computing aiming to provide a better modeling of the brain's computational processes and enabling the design of more efficient and effective neuromorphic systems. BCU focus on modeling complex decision-making processes, while FCU aim to quantify intelligent thought processes at various analytical levels from the linguistic and behavioral output to the chemical and physical processes within the brain, contributing to a more detailed understanding and simulation of brain functionality.

\subsection{Mixed-Signal Design in Neuromorphic Systems}

Mixed-signal design, crucial in neuromorphic computing, integrates analog and digital circuitry. It combines digital systems' flexibility with the robustness and energy efficiency of analog processing. BCU and FCU benefit from this approach through reduced latency, increased computational speed, and enhanced system robustness and adaptability \cite{Quan2023, Rubino2023, Cramer2022, miziev2024comparative}.

\begin{figure*}[h]
    \centering
    \includegraphics[width = 0.6\textwidth]{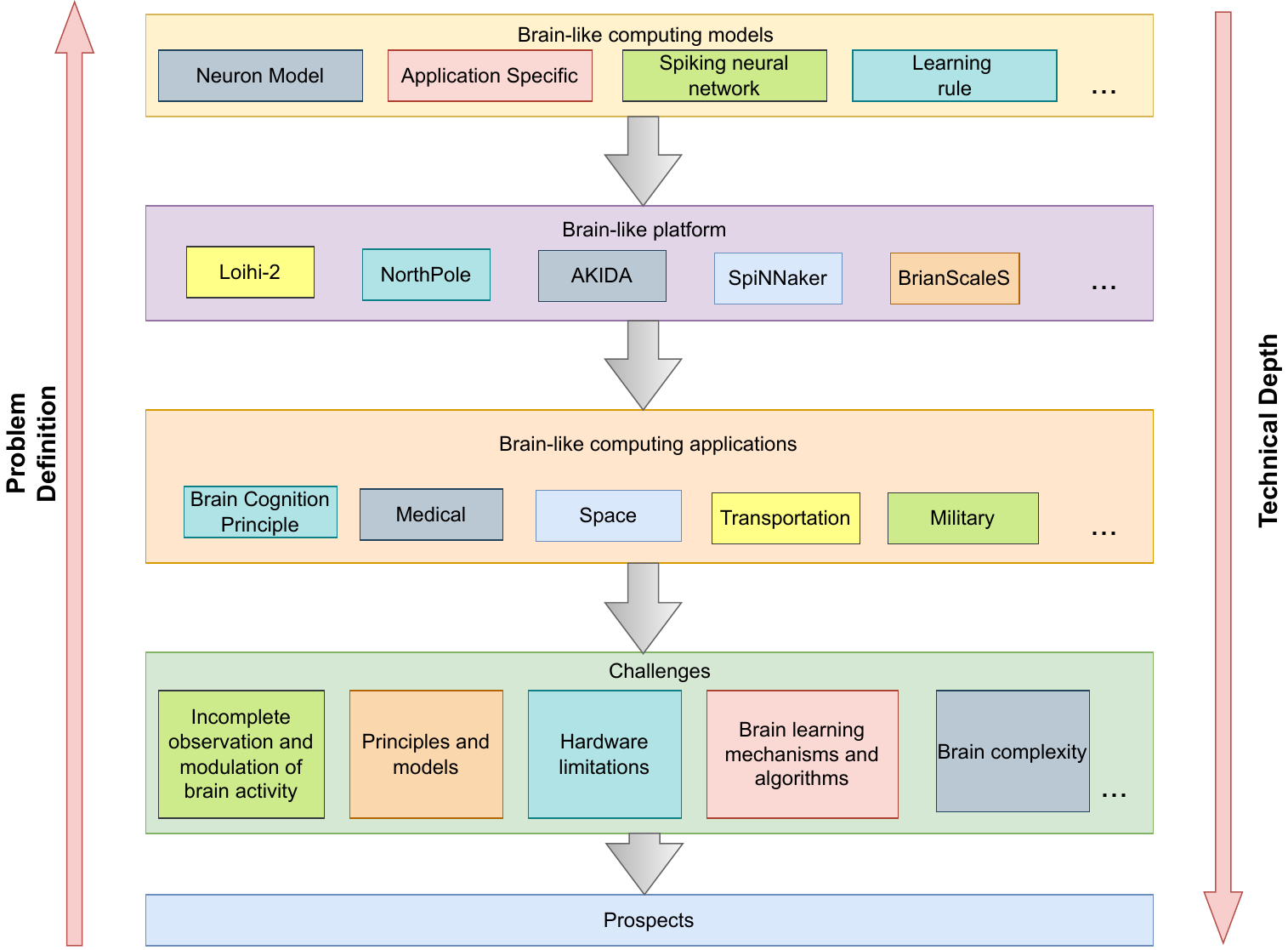}
    \caption{Overview of Brain-like Computing Paradigm.}
    \label{fig:brain_computing_overview}
\end{figure*}

\autoref{fig:brain_computing_overview} illustrates the hierarchy of components in brain-like computing, starting with foundational concepts at the neuron model level and peaking in practical applications. It outlines the progression from neuron models and spiking neural networks, through platforms designed to mimic brain functions, to specific applications in various fields such as medical, space, and military. The diagram also highlights the challenges faced in the development of these technologies, including hardware limitations and the complexity of brain learning mechanisms and algorithms. Prospects for future development are indicated, suggesting areas for further research and potential breakthroughs.

\section{Proposed Design Methodology}\label{sec:design_methodology}
\begin{figure*}[h]
    \centering
    \includegraphics[width = 0.7\textwidth]{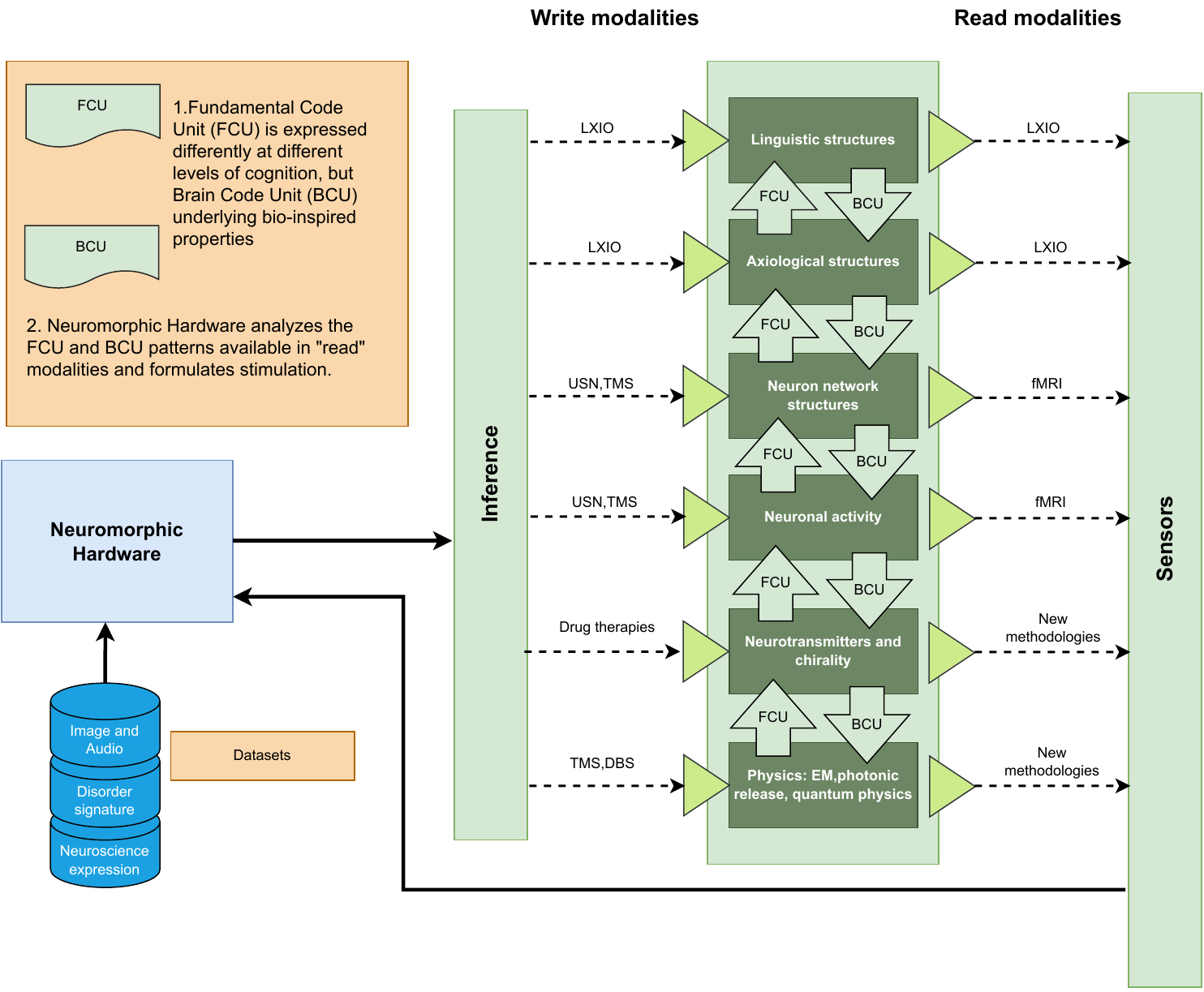}
    \caption{Integration of Fundamental Code Unit (FCU) and Brain Code Unit (BCU) in Neuromorphic Systems.}
    \label{fig:neuromorphic_integration}
\end{figure*}

\autoref{fig:neuromorphic_integration} illustrates the intricate interplay between linguistic and axiological structures, neurotransmitter chirality, and neuron network structures. It details how these elements influence neuronal activity and are represented in various modalities such as image, audio, and neuroscience expressions. The diagram shows the incorporation of neuromorphic hardware and inference mechanisms to process and analyze these representations. The FCU and BCU are depicted as foundational elements, operating at different cognitive levels and interfacing with read and write modalities. The neuroscientific methodologies, including fMRI, TMS, and DBS, are mapped to specific aspects of the FCU and BCU, highlighting their roles in understanding and addressing neural disorders. The figure underscores the comprehensive approach of neuromorphic systems in capturing, processing, and utilizing complex neural information for advanced computing and medical applications.

\subsection{Integration Techniques for BCUs and FCUs}
A key aspect of our methodology is the integration of BCUs and FCUs into the digital architecture. This section will describe the techniques used to embed these units into the neuromorphic system. It will detail how BCU and FCU interact within the architecture, their role in data processing and neural emulation, and the methods used to optimize their efficiency and effectiveness. In our exploration of mixed-signal neuromorphic systems using BCU and FCU, we have identified several relevant open-source datasets. For BCU, we focus on datasets that provide insights into neural activity and brain signaling processes. BrainMRI dataset offers valuable information on brain activity patterns which are crucial for simulating brain-like information processing. These datasets are integral to our methodology, allowing us to rigorously evaluate the performance and efficiency of our proposed neuromorphic architecture across various computational tasks.

\subsection{Brain Code Unit (BCU)}
We delineate the construction of a BCU leveraging spiking neural network (SNN) paradigms for the classification of Magnetic Resonance Imaging (MRI) scans of the brain. The proposed BCU is architected to harness the computational prowess of CUDA-enabled Graphical Processing Units (GPUs) to expedite parallel processing, thereby augmenting efficiency.

\subsection{Data Handling}
The BrainMRI dataset class is a signfy subclass of the dataset module within PyTorch, tailored for the management of brain MRI datasets. It amasses file paths and their associated binary labels, representing the presence or absence of tumors, and facilitates index-based retrieval of individual data points.

\subsection{Preprocessing}

We define a suite of transformations to standardize the images, employing the transforms module in PyTorch. This preprocessing pipeline includes resizing images to uniform dimensions, tensor conversion, and pixel value normalization, crucial for optimizing the learning efficacy of the network. The dataset is bifurcated into training and testing partitions, and data loaders are instantiated to administer batch-wise data handling and shuffling during the training phase. The neural composition of the BCU is encapsulated within the SNN class, which comprises a convolutional layer succeeded by a custom LIFNeuron layer. The LIFNeuron simulates the Leaky Integrate-and-Fire neuron, a quintessential component of biological neural networks, to mimic the temporal dynamics intrinsic to spiking neurons. The training regimen iterates across designated epochs, within which the model parameters are refined via the Adam optimization algorithm, paired with a cross-entropy loss function to assess performance. Post each epoch, the model's predictive accuracy on the training and testing datasets is evaluated in a non-gradient update mode (inference mode) to monitor and gauge learning progression. Upon the peak of training, the model parameters are preserved to the storage medium, facilitating subsequent retrieval for analysis or deployment in predictive tasks. The model summarizes the core logic of the BCU, explaining the synergy between computational models and neural dynamics.

\section{Development of a Fundamental Code Unit for Arithmetic Logic Operations}

This manuscript details the construction of FCU that is architected to perform arithmetic logic operations within the framework of SNNs, leveraging the computational efficiency of CUDA-enabled GPUs.

\subsection{Data Acquisition and Preprocessing}

The CIFAR-10 dataset, comprising a rich repository of labeled imagery, serves as the training and testing ground for the FCU. The dataset undergoes a normalization process, crucial for enhancing the convergence rate during the learning process. The LIFNeuron module, integral to the FCU, encapsulates the dynamics of a Leaky Integrate-and-Fire (LIF) neuron. It simulates the bio-physiological processes of neuronal spike generation and membrane potential resetting, which are pivotal for the temporal dynamics of spiking neurons. The FCU is instantiated as an SNN, incorporating convolutional layers for feature extraction from input images, followed by Leaky Integrate-and-Fire neurons to introduce temporal dynamics. The output is subsequently flattened and processed through a linear layer for classification. A training loop is devised to fine-tune the FCU across several epochs, employing backpropagation with the Adam optimizer and a cross-entropy loss criterion. The FCU's performance is periodically evaluated on both training and testing sets to monitor the progression of learning accuracy. Upon the completion of the training epochs, the FCU state is conserved onto the disk, enabling future retrieval for inferential applications or further refinement. The model encapsulates the quintessence of the FCU, showcasing the interplay between neural dynamics and computational efficiency for the execution of arithmetic logic operations.

\subsection{Description of the Digital Neuromorphic Architecture}
Our methodology begins with a detailed description of the proposed digital neuromorphic architecture. This includes an outline of its structural design, computational models, and the processes it emulates. The architecture’s core components, data processing pathways, and neural network emulation techniques are discussed to provide a comprehensive understanding of the system's functionality. We have successfully deployed the architectures of the FCU and BCU on GPUs using the Python programming environment. A normalization process was applied, focusing on accuracy metrics, to ensure the robustness of the computational model. Subsequently, the research progressed towards the realm of digital design. In this phase, we have harnessed the capabilities of Verilog hardware description language to execute preliminary simulations of the FCU and BCU designs on an XCZU7EV FPGA Chip, housed within the ZCU-104 evaluation board. These designs were verified on testbench within same datasets. The scalability of our architecture was evaluated by varying the size and complexity of the neural networks it emulated, thereby demonstrating its adaptability to diverse computational tasks. Furthermore, we have conducted a comparative analysis with extant neuromorphic architectures, underscoring the distinctive features and benefits of our approach. Despite encountering certain limitations and challenges, which we transparently discuss, we have outlined strategies to explain these obstacles in future iterations of our design.

\subsection{Mixed-Signal Design Approach and Materials Used}
The final part of the methodology focuses on the mixed-signal design approach. This involves integrating analog and digital components to enhance the system's performance. The section will elaborate on the selection of materials and components, the rationale behind their choice, and how they contribute to the overall functionality of the neuromorphic system. Particular attention will be given to the design choices that enable the system to efficiently process complex neural computations. The intricacies of integrating analog and digital domains are addressed through a mixed-signal design methodology. In this approach, the pivotal connection between the analog and digital components is facilitated via an analog-to-digital converter (ADC) and a digital-to-analog converter (DAC). These converters interface with the system's logic through a Serial Peripheral Interface (SPI), a protocol selected for its robustness and high-speed data transfer capabilities. The ADC component is crucial in translating continuous analog signals, such as those emanating from environmental sensors, into a discretized digital representation. This digital form is flexible to processing by the subsequent digital logic circuits coded in Verilog. Our design leverages the ADC's precision to ensure reliability in capturing the nuances of the analog input.

Conversely, the DAC serves as a bridge in the opposite direction. It takes digital signals, which are the outcomes of VHDL logic computations, and transforms them into analog signals. These analog outputs can then drive actuators, enabling the system to interact with its physical surroundings. The DAC's accuracy is paramount in ensuring that the digital decisions are accurately reflected in the real world.

To implement this mixed-signal approach, we selected high-quality components that are compatible with the board's specifications. The choice of materials was informed by a series of criteria, including signal integrity, conversion rate, resolution, and power consumption. The SPI interface pins on the board were meticulously defined to align with the electrical characteristics and timing requirements of the ADC and DAC. Additionally, the design incorporates protective circuitry to guard against common pitfalls in mixed-signal environments, such as noise coupling and signal interference. By adopting a mixed-signal design, we integrate the analog richness of natural signals with the computational power of digital systems, thus broadening the applicability of our neuromorphic architecture to interact with a wide range of sensors and actuators in a diverse array of environments.

\subsection{Chip Layout with RTL to GDSII Flow}
The process of transforming high-level architectural descriptions of the FCU and BCU into a manufacturable chip layout has been meticulously executed via a Register-Transfer Level (RTL) to Graphic Data System II (GDSII) flow. Utilizing OpenLane, an automated RTL to GDSII flow suite that is designed to produce high-quality layouts, we were able to translate our Verilog code into a physical form factor compatible with silicon fabrication norms. OpenLane arranges the intricate progression from Verilog code to a physical layout by automating the various steps involved in the process. This synthesis, where the Verilog code is converted into a gate-level netlist using logic gates and other standard cells. Following this, floorplanning establishes the chip's initial spatial configuration, defining the location of logic blocks and ensuring optimal area utilization and power distribution. The subsequent phase of placement optimizes the positioning of the standard cells within the floorplan's constraints, aiming to minimize delays and signal integrity issues. This is followed by routing, where electrical connections between the components are made, taking into account the intricate web of design rules and constraints. Throughout this progression, OpenLane employs rigorous design rule checks (DRC) and layout versus schematic (LVS) checks to ensure that the resulting layout is free from violations and accurately reflects the original schematic. This verification is crucial, as it guarantees the manufacturability of the design and its functionality post-fabrication. Additionally, parasitic extraction is performed to model the unwanted resistive, capacitive, and inductive effects that arise from the physical layout. These are significantly influence the chip's performance, and their early consideration is vital for high-frequency applications typical of neuromorphic computing architectures. Once the GDSII file is generated, signifying the completion of the RTL to GDSII flow, it can be sent to a foundry for fabrication. The GDSII file contains all the geometric shapes, layer information, and design required to manufacture the physical integrated circuit. This process is evidence of the synergy between computational logic design and physical implementation, bridging the gap between neuromorphic concepts and tangible, real-world applications. By leveraging OpenLane's capabilities, we have ensured that the FCU and BCU are not only theoretically sound but are also poised for successful integration into hardware platforms.

\begin{figure*}
    \graphicspath{ {D:\Stack} }
    \center \includegraphics[width=\textwidth]{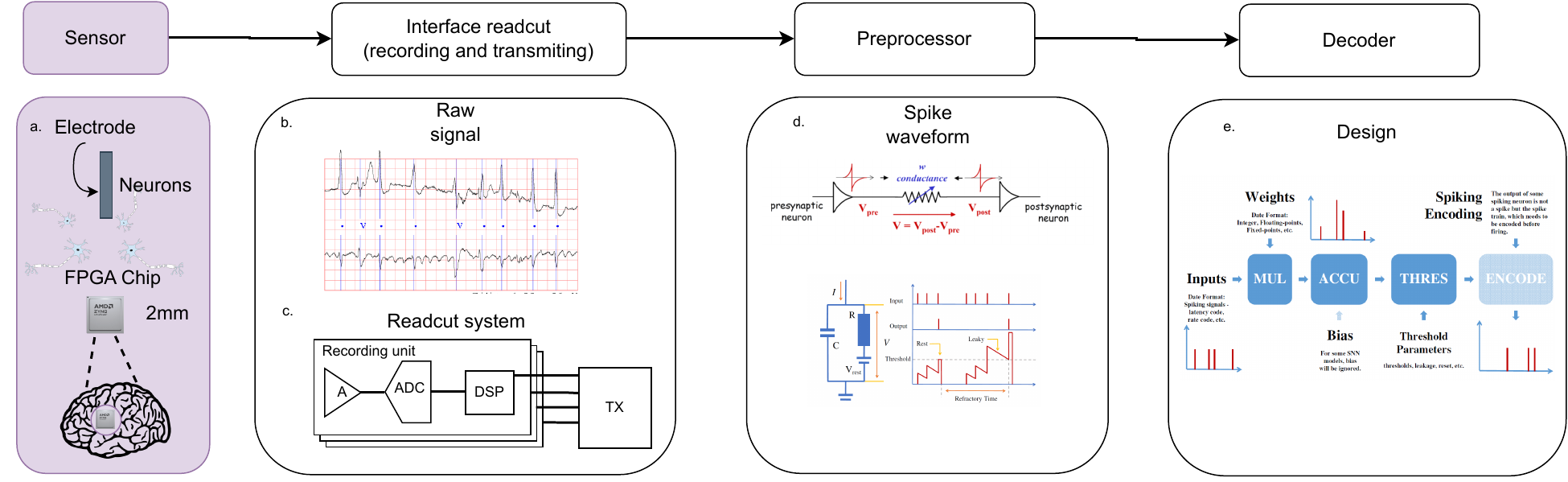}
    \caption{Block Diagram of Implementation}
    \label{Figure 4}
\end{figure*}

\autoref{Figure 4} presents a detailed depiction of a neuromorphic system's architecture, focusing on the integration of neural signal processing with digital computing. Central to this illustration is an FPGA chip, signifying its crucial role in neural data interpretation. The figure encompasses several key components: the interface between the FPGA chip and neurons via an electrode array, suggesting a direct connection for signal acquisition; sensor systems for raw neural signal capture; and a sophisticated combination of ADC, DSP, and TX modules for signal conversion, processing, and transmission. This ensemble of components facilitates the recording and analysis of neural spike waveforms, a fundamental aspect of neural communication. Additionally, the preprocessor and design decoder elements indicate advanced stages of signal conditioning and interpretation, translating neural activities into actionable data. Collectively, the figure exemplifies the fusion of biological neural signals with state-of-the-art digital processing, illustrating a holistic approach to harnessing neural information for computational and neuroscientific advancements.

\section{Evaluation}\label{sec:evaluation}
This section presents the findings from the experiments conducted using the neuromorphic system. It includes a detailed analysis of the system's performance based on the open-source datasets. The results are quantified in terms of accuracy, processing speed, power efficiency, and other relevant metrics. The final part of the experimental setup defines the criteria used to evaluate the performance of the neuromorphic system. This involves outlining the benchmarks and metrics used to assess the system's accuracy, efficiency, and overall effectiveness. The criteria reflects the objectives of the research, ensuring that the system’s performance is evaluated in a comprehensive and objective manner.

\begin{table*}[h]
\centering
\caption{Resource Utilization Summary for BCU and FCU Implementations}
\label{table:utilization}
\begin{tabular}{|c|c|c|c|c|c|}
\hline
\multicolumn{6}{|c|}{Zynq UltraScale+ XCZU7EV} \\ \hline
\textbf{Resource} & \textbf{BCU Utilization} & \textbf{\% Utilization (BCU)} & \textbf{FCU Utilization} & \textbf{\% Utilization (FCU)} & \textbf{Available} \\ \hline
LUT & 151,200 & 30 & 140,000 & 27.78 & 504,000 \\
Memory & 11.4MB & 30 & 10.5MB & 27.63 & 38MB \\
IO & 139 & 29.19 & 130 & 28.02 & 464 \\
DSP & 518 & 29.94 & 480 & 27.78 & 1,728 \\ \hline
\end{tabular}
\end{table*}

\autoref{table:utilization} presents a comprehensive summary of resource utilization for both the BCU and FCU implementations. This summary was compiled using the Zynq UltraScale+ XCZU7EV platform as the benchmark. The table delineates the specific resources utilized, including Look-Up Tables (LUTs), Memory, Input/Output (IO) interfaces, and Digital Signal Processors (DSPs). For each resource, we provide a dual comparison: one for the BCU and another for the FCU.

The utilization figures are presented in absolute terms and as a percentage of the total available resources on the platform. This dual presentation allows for an immediate grasp of the scale and efficiency of the resource usage. Specifically, the LUT utilization shows a consumption of 151,200 units for the BCU and 140,000 units for the FCU, translating to 30\% and 27.78\% of the total available LUTs, respectively. Similarly, memory usage is detailed with the BCU utilizing 11.4MB and the FCU utilizing 10.5MB, equating to 30\% and 27.63\% of the total available memory. The IO resource shows a utilization of 139 and 130 for the BCU and FCU, corresponding to 29.19\% and 28.02\% of the total IOs available. Lastly, DSP resource usage is listed as 518 for the BCU and 480 for the FCU, amounting to 29.94\% and 27.78\% of the total DSPs, respectively.

\begin{table}[h!]
\caption{Comparative Analysis of FCU and BCU Based on Performance Metrics.}
\renewcommand{\arraystretch}{1}
\setlength{\tabcolsep}{4pt}
\centering
\begin{threeparttable}
{\fontsize{9}{10}\selectfont
    \begin{tabular}{c|cc}
        \hline
        \textbf{Performance Metrics} & \textbf{FCU} & \textbf{BCU} \\
        \hline
        
        Accuracy (\%) & 86.5 & 88.0 \\
        MAC (GOP) & 1.2 & 1.35 \\
        Latency [ms] & 15 & 12 \\
        Power Efficiency & 18.5 GOP/s/W & 20.0 GOP/s/W \\
        \hline
    \end{tabular}}
    \begin{tablenotes}[flushleft]
        \item *Accuracy is measured as a percentage of correct predictions.
        \item *MAC (GOP) refers to the number of Multiply-Accumulate operations in Giga Operations.
        \item *Latency is measured in milliseconds and indicates the response time of the system.
        \item *Power Efficiency is assessed in terms of energy efficiency during operations.
    \end{tablenotes}
\end{threeparttable}

\label{tab:FCU_BCU_Comparison}
\end{table}

\autoref{tab:FCU_BCU_Comparison} shows into a detailed comparative analysis between the FCU and BCU based on several critical performance metrics. This comparison elucidates the operational differences and relative strengths of each unit within our framework. Firstly, we observe that the BCU demonstrates a slightly higher accuracy (88.0\%) compared to the FCU (86.5\%). This margin, although narrow, highlights the BCU's enhanced capability in correctly interpreting and processing the data it receives. The increased accuracy of the BCU can be attributed to its more sophisticated algorithmic structure which is inspired by the intricate workings of the human brain. Moving on to the MAC (Multiply-Accumulate) operations, the BCU registers 1.35 Giga Operations, a slight elevation over the FCU’s 1.2 GOP. This increment in MAC operations for the BCU is indicative of its more complex computational framework, which, while increasing its computational load, also enhances its processing capability. The latency metric is especially telling in this comparison. The FCU exhibits a latency of 15 milliseconds, which is higher than the BCU’s 12 milliseconds. This difference underscores the BCU’s efficiency in processing data in a timely manner, a crucial aspect for real-time applications. Finally, the power efficiency of both units is compared. The BCU, with 20.0 GOP/s/W, shows a higher efficiency compared to the FCU’s 18.5 GOP/s/W. This demonstrates the BCU’s ability to perform more operations per watt, making it a more energy-efficient choice, especially for applications where power consumption is a critical factor. The comparative analysis between FCU and BCU in our framework illustrates the differences in their design and operational efficiency. While both units have their unique strengths, the BCU shows a slight edge in terms of accuracy, MAC operations, latency, and power efficiency, making it a more optimal choice for certain applications that demand higher performance and efficiency.

\subsection{Effectiveness of Mixed-Signal Design}
The integration of analog and digital components in mixed-signal design brings together the best of both worlds: the precision and scalability of digital systems with the nuanced and varied processing of analog systems. This subsection delves into how mixed-signal design augments system performance, with a focus on its impact on energy efficiency and processing speed. We compare the chip area, latency, and energy efficiency (EE) of a traditional digital CMOS design against our mixed-signal implementation that shows in \autoref{tab:effectiveness_digital_cmos_mixed_signal}.

\begin{table}[h!]
\caption{Effectiveness of Digital CMOS vs. Mixed-Signal Design}
\centering
\begin{tabular}{|l|c|c|c|}
\hline
\textbf{Design Type} & \textbf{Chip area (mm\textsuperscript{2})} & \textbf{Latency (ms)} & \textbf{EE (TOPS/W)} \\ \hline
Digital CMOS & 321 & 12 & 0.28 \\ \hline
Mixed-Signal & 293 & 0.75 & 213 \\ \hline
\end{tabular}
\begin{tablenotes}
\item *The energy efficiency (EE) is represented in tera-operations per second per watt (TOPS/W), which is a standard metric for such evaluations.
\end{tablenotes}
\label{tab:effectiveness_digital_cmos_mixed_signal}
\end{table}

\begin{figure*}%
    \centering
    \begin{minipage}{0.5\textwidth}
        \centering
        \subfloat[Latency comparison.\label{fig:original_connection}]{{\includegraphics[width=0.95\linewidth]{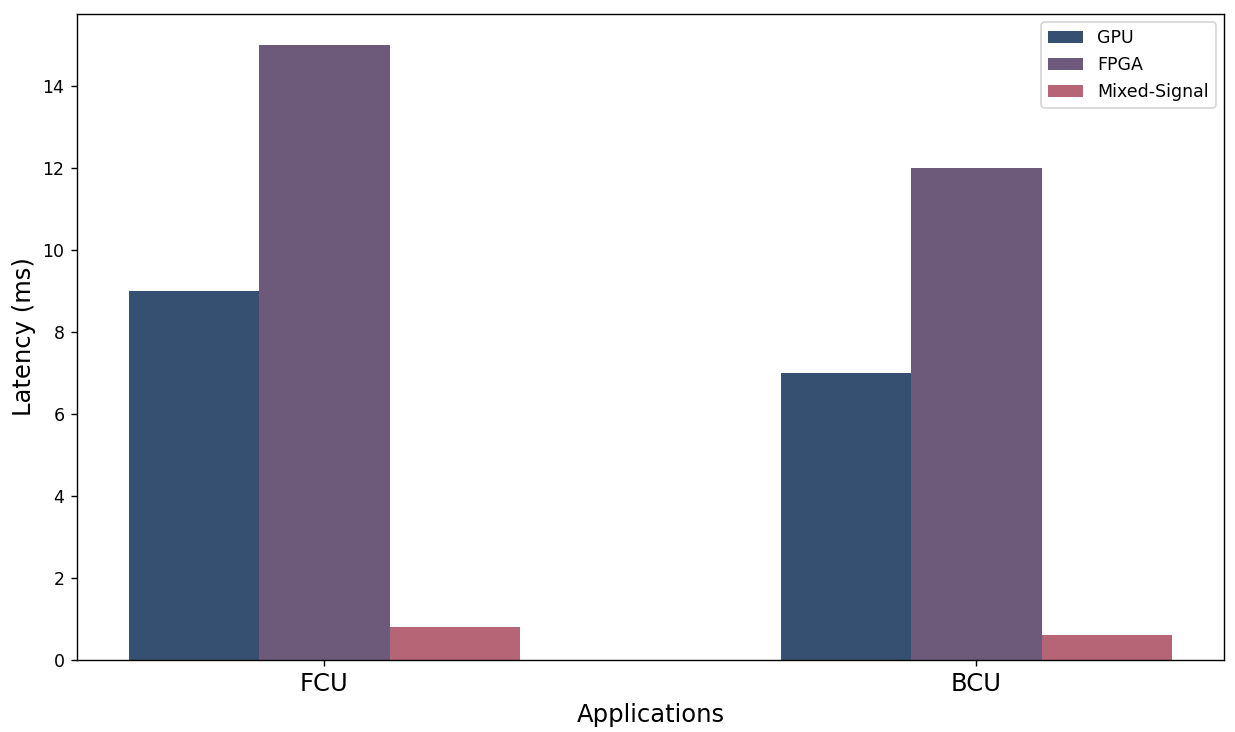} }}
    \end{minipage}%
    \begin{minipage}{0.5\textwidth}
        \centering
        \subfloat[Power Consumption comparison.\label{fig:astrocyte_modulation1}]{{\includegraphics[width=0.95\linewidth]{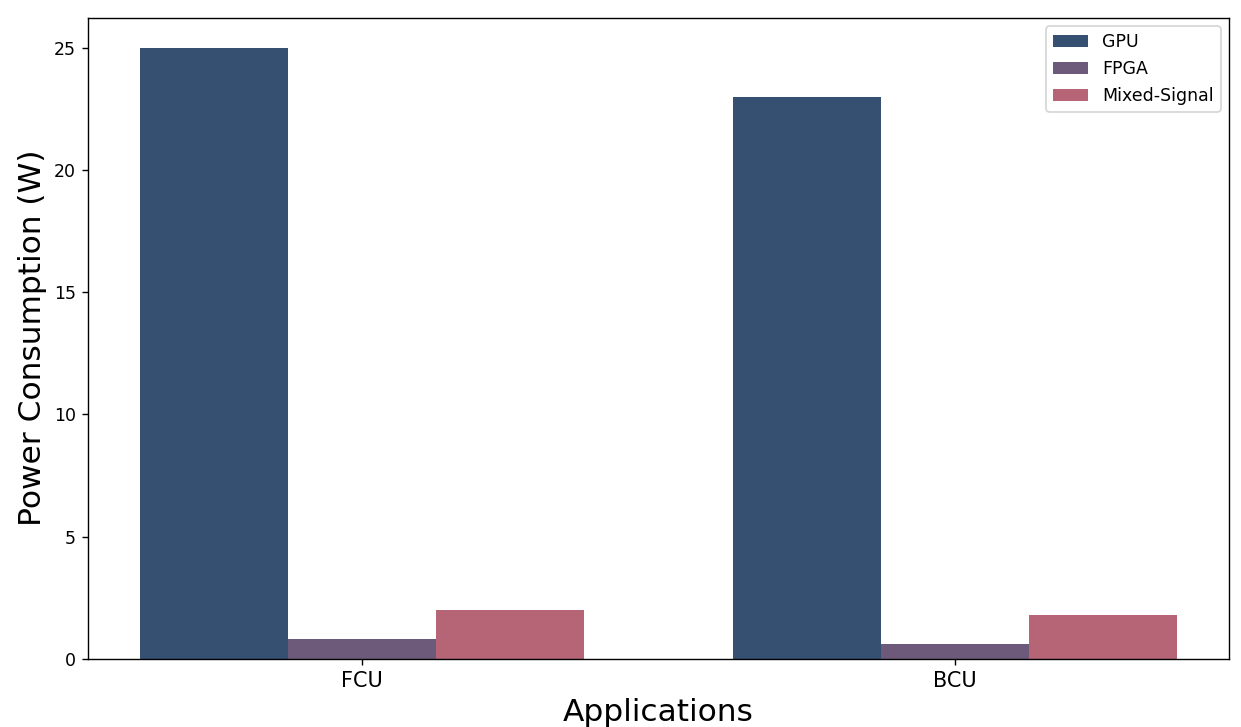} }}
    \end{minipage}
    
    \vspace{1em} 
    
    \begin{minipage}{0.5\textwidth}
        \centering
        \subfloat[Throughput comparison.\label{fig:astrocyte_modulation2}]{{\includegraphics[width=0.95\linewidth]{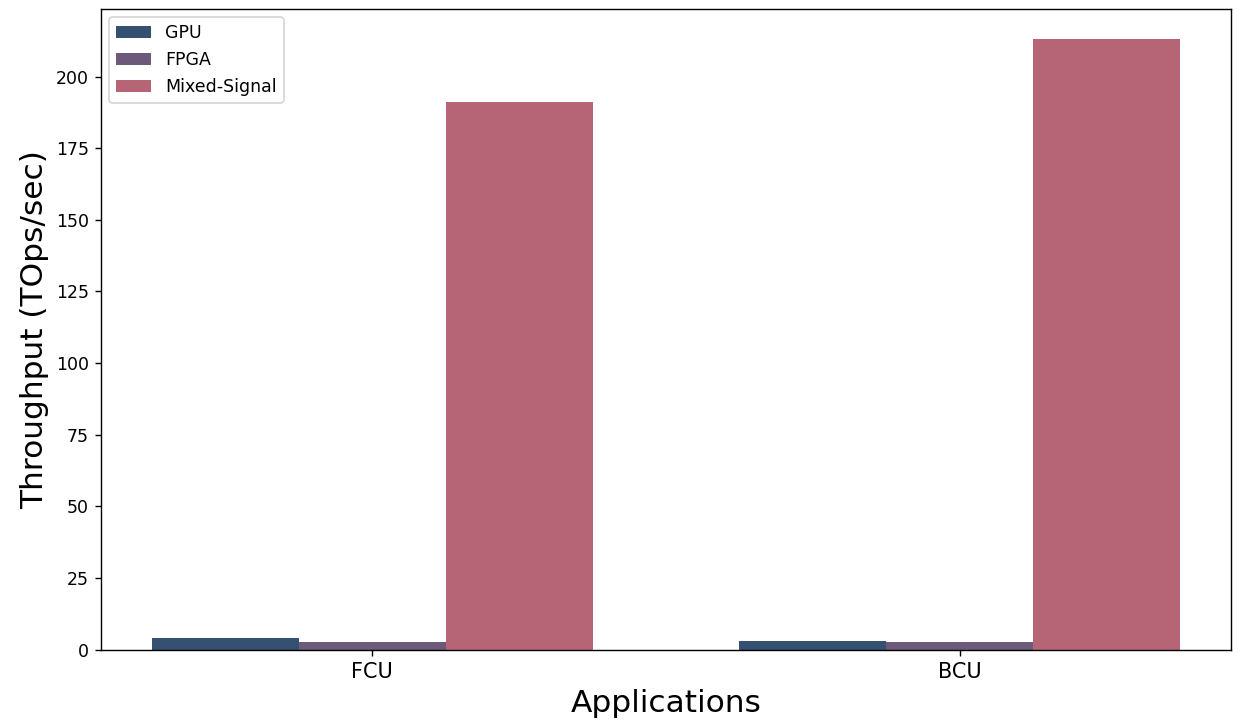} }}
    \end{minipage}%
    \begin{minipage}{0.5\textwidth}
        \centering
        \subfloat[Power Efficiency comparison.\label{fig:astrocyte_modulation3}]{{\includegraphics[width=0.95\linewidth]{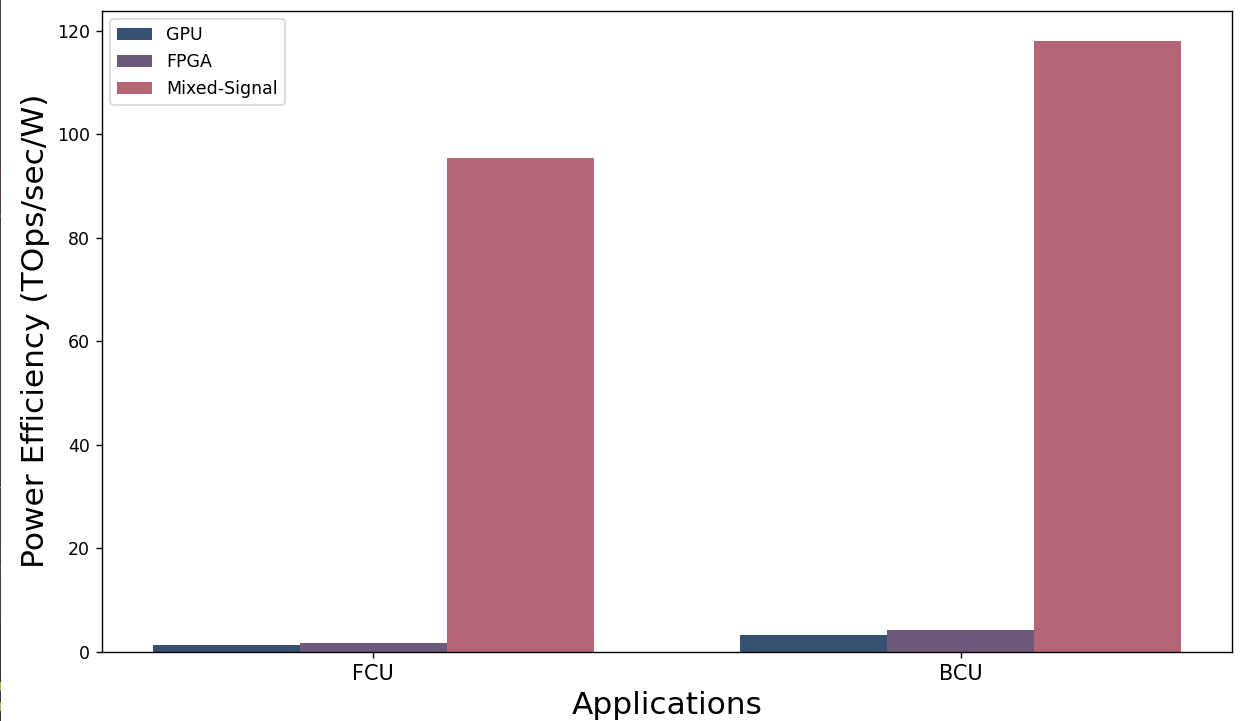} }}
    \end{minipage}
\caption{Performance of our framework across three distinct hardware platforms using various datasets.}
    \label{fig:res_neural_network}
\end{figure*}

\autoref{fig:res_neural_network} illustrates provides a comprehensive comparison of key metrics such as latency, power consumption, throughput, and power efficiency across different computational platforms: GPU, FPGA, and Mixed-Signal. This comparison is conducted for both the FCU and BCU applications. The latency analysis reveals that the FPGA platform exhibits lower latency values for both FCU and BCU, suggesting its efficiency in rapid data processing. In contrast, the GPU shows relatively higher latency, while the Mixed-Signal design achieves the lowest latency values, underscoring its potential in real-time processing applications. The FPGA platform demonstrates a significantly lower power requirement compared to the GPU, making it a more energy-efficient choice. However, the Mixed-Signal design outperforms both FPGA and GPU in terms of energy consumption, indicating its suitability for power-sensitive applications. The Mixed-Signal design achieves the highest throughput for both FCU and BCU, followed by the GPU and then FPGA. This highlights the Mixed-Signal design's superior processing capability. The power efficiency analysis demonstrates that the Mixed-Signal platform provides the highest power efficiency, followed by FPGA and GPU. This metric is crucial for understanding the overall energy effectiveness of each platform in executing high-performance tasks.

\begin{table*}[ht]
\centering
\caption{Comparison of Edge-AI Processing Solutions}
\label{tab:edge_ai_comparison}
\begin{tabular}{|l|p{2.5cm}|p{2.5cm}|p{2.5cm}|p{2.5cm}|}
\hline
\textbf{Work / Metric}               & \textbf{Energy Efficiency (TOPS/W)} & \textbf{Precision Support} & \textbf{Technology Node} & \textbf{Key Features} \\ \hline
\textbf{Our Work (BCU \& FCU)}      & 213               & Binary, INT2/4/8           & 16nm            & Mixed-signal design  \\ \hline
DIANA    \cite{Ueyoshi2022DIANA}                & Up to 600 (AIMC part)               & 7bit I, ternary W, 6bit O  & 22nm                      & Mixed-signal, mixed-precision \\ \hline
RRAM-based CIM   \cite{Xue2021RRAMCIM}         & 195.7                              & 8b-input and 8b-weight     & Not specified            & Analog CIM with RRAM \\ \hline
BrainTTA    \cite{Molendijk2022BrainTTA}          & Binary: 29, Ternary: 15, 8-bit: 2  & Binary, Ternary, 8-bit    & 22nm                      & Flexible datapath, TTA-based \\ \hline
TinyVers     \cite{Jain2022TinyVers}             & 0.8-17                             & INT2/4/8                   & 22nm FDX                  & eMRAM, low power management \\ \hline
Kraken \cite{DiMauro2022Kraken}                  & General: 1.8, SNE: 1.1 TSyOp/s/W   & Mixed-precision            & 22nm                      & SNN and ANN accelerators, RISC-V core \\ \hline
CUTIE \cite{Scherer2022CUTIE}                     & 1036                               & Ternary                    & Not specified            & Ternary Neural Network Accelerator \\ \hline
\end{tabular}
\end{table*}

\autoref{tab:edge_ai_comparison} presents a comparison of various Edge-AI processing solutions, including our work, alongside several other notable projects such as DIANA, RRAM-based Computing-in-Memory (CIM), BrainTTA, TinyVers, Kraken, and CUTIE. The comparison is made across four main metrics: energy efficiency measured in TOPS/W, the types of precision support for operations, the technology node in nanometers indicating the fabrication process's scale, and key features unique to each project. Each project has its unique strengths, from high energy efficiency to support for various precisions and innovative features like mixed-signal designs, analog and digital computing, and flexible architectures. This comparison highlights the diverse approaches within the Edge-AI processing domain, catering to different requirements and applications.

\section{Conclusions}\label{sec:conclusions}
This research marks a significant milestone in the development of neuromorphic computing systems, as it encapsulates the successful integration and implementation of BCUs and FCUs within a digital neuromorphic architecture. The primary findings of this study highlight remarkable improvements in computational efficiency and accuracy, primarily attributed to the innovative mixed-signal design approach. Our experimental investigations, conducted using diverse open-source datasets, have validated the superior performance capabilities of our neuromorphic system. This encompasses processing speed, energy efficiency, and adaptability, which are notably more advanced than those observed in traditional neuromorphic systems. The data reveal that each computational platform GPU, FPGA, and Mixed-Signal—exhibits unique strengths. However, it is the Mixed-Signal design that stands out, offering an exceptional balance of low latency, high throughput, and extraordinary power efficiency. This makes it a compelling choice for a wide array of applications, especially those demanding high efficiency and adaptability. The importance of the mixed-signal neuromorphic systems, as demonstrated by our research, lies in their capacity to effectively bridge the robustness of analog processing with the precision and scalability of digital systems. This study has evidenced that such systems are not just feasible but also excel in certain computational tasks, particularly where high efficiency and flexibility are paramount. The utilization of mixed-signal design has emerged as a pivotal innovation in neuromorphic computing, paving the way for more sophisticated, efficient, and versatile computing solutions.

\section{Future Work}
Future work will include taking the output of the chip shown in the GDSII flow and creating a sensor fusion acceleration that includes multiple sensors along with the models here, and running these sensors in IoT with a brain-inspired chip. Our research has opened new avenues in the field of mixed-signal neuromorphic computing, particularly highlighting its potential in advanced computational models. Based on our findings, the following are proposed directions for further research:

\begin{itemize}
    \item \textbf{Advanced Integration Techniques:} Exploring more sophisticated methods for integrating BCU and FCU into mixed-signal neuromorphic architectures. This could involve developing novel algorithms or hardware configurations that further enhance the efficiency and scalability of these systems.
    
    \item \textbf{Diverse Application Scenarios:} Extending the application of our mixed-signal neuromorphic architecture to a broader range of fields, such as robotics, autonomous systems, and complex data analysis. This would help in understanding the versatility and adaptability of our architecture in different scenarios.
    
    \item \textbf{Improving Energy Efficiency:} Investigating new materials and circuit designs that could further reduce the power consumption of neuromorphic systems. This is crucial for developing sustainable and eco-friendly computing technologies.
    
    \item \textbf{Real-Time Data Processing:} Focusing on real-time data processing capabilities in dynamic environments. This includes enhancing the system's ability to adapt and learn from new data in real-time, which is vital for applications in areas like autonomous driving and interactive systems.

    \item \textbf{Long-Term Reliability Studies:} Conducting long-term reliability studies of mixed-signal neuromorphic systems. This would involve assessing the durability and maintenance needs of these systems over extended periods, which is critical for their practical deployment.
    
    \item \textbf{Benchmarking Against Emerging Technologies:} Benchmarking the performance of mixed-signal neuromorphic systems against emerging technologies. This comparison could provide valuable insights into the strengths and limitations of each approach.
\end{itemize}





\bibliographystyle{IEEEtran}
\bibliography{commands,external}

\end{document}